# Very long baseline interferometry and observations of gravitational lenses using intensity fluctuations: an analysis based on intensity autocorrelation

Ermanno F. Borra,
Centre d'Optique, Photonique et Laser,
Département de Physique, Université Laval, Québec, Qc, Canada G1K 7P4
(email: borra@phy.ulaval.ca)

(SEND ALL EDITORIAL CORRESPONDENCE TO : E.F. BORRA)

RECEIVED________________________________________

# ABSTRACT

A novel interferometric technique that uses the spectrum of the current fluctuations of a quadratic detector, a type of detector commonly used in Astronomy, has recently been introduced. It has major advantages with respect to classical interferometry. It can be used to observe gravitational lenses that cannot be detected with standard techniques. It can be used to carry out very long baseline interferometry. Although the original theoretical analysis, that uses wave interaction effects, is rigorous, it is not easy to understand. The present article therefore carries out a simpler analysis, using the autocorrelation of intensity fluctuations, which is easier to understand. It is based on published experiments that were carried out to validate the original theory. The autocorrelation analysis also validates simple numerical techniques, based on the autocorrelation, to model the angular intensity distribution of a source. The autocorrelation technique also allows a much simpler detection of the signal.

In practice, the gravitational lens applications are the ones that can readily be done with presently available telescopes. We describe a practical example that shows that presently available VLBI radio-astronomical data can be used to observe microlensisng and millilensing in macrolensed Quasars. They may give information on the dark matter substructures in the lensing galaxies.

# 1. INTRODUCTION

Most astronomical interferometers use amplitude interferometry. Intensity interferometry (Hanbury-Brown 1968) is used less frequently. Borra (2008) proposes a novel technique that uses the spectrum of intensity fluctuations to obtain the time delays in gravitational lenses. The analysis in Borra (2008) is based on the standard model for gravitational lenses, which is a Young interferometer (Refsdal 1964, Press & Gunn 1973). Borra (2013) shows that the technique can also be applied in long baseline astronomical interferometry. The theoretical analysis in Borra (2008) is experimentally confirmed by Borra (2011).

This novel type of interferometer has advantages over classical interferometry, but also disadvantages. Its outstanding advantage comes from the fact that the angular intensity distribution of a source can be determined by measuring how the visibility of the spectral modulation of the spectrum of intensity fluctuations of the output current of a quadratic detector varies as a function of its frequency, while keeping the separation between 2 telescopes (baseline) constant. The frequency spectrum is easily obtained with software that first computes the autocorrelation of the intensity as a function of time and then carries out the Fourier transform of the autocorrelation. In amplitude and intensity interferometers, the angular intensity distribution of a source is obtained by changing the separation between telescopes or using several telescopes located at different constant baselines. In our case only 2 telescopes at fixed locations are needed. Another major advantage is that it can use extremely large baselines, thereby allowing extremely small angular resolutions. Its major disadvantage comes from the fact that one needs a greater separation between telescopes than classical interferometry to obtain the same angular resolution. In practice the technique will allow much greater resolutions because of the extremely large baselines that it can use. For example section 2 below discusses the observation of a binary star by two optical telescopes separated by 1000 km. The advantages and disadvantages are discussed in Borra (2013) for optical and radio-telescopes.

For the study of gravitational lenses, the major advantages of the technique are: Firstly, that very short time delays (e.g. $10^{-7}$ seconds) as well as long time delays (e.g. hundreds of seconds) can be detected in a single observation lasting an hour. Secondly the detection of the time-delay signature unambiguously identifies the lensing event. The techniques currently used necessitate observations of luminosity variations over very long times (weeks or months).

Borra (2008) follows a rigorous mathematical treatment based on the superposition of electromagnetic waves, which is the correct physical theory that directly comes from Maxwell's equations. Consequently, there is little doubt that the theoretical analysis in Borra (2008) is valid, particularly considering that it is validated by laboratory work (Borra 2011). However, the treatment in Borra (2008) is difficult to understand for readers unfamiliar with statistical optics. The present article therefore gives a far easier to follow treatment that uses intensity autocorrelation. Another major advantage of using the autocorrelation is that it is far more intuitive than the Fourier transform analysis of the spatial coherence function used in Borra (2008) to analyze the data and therefore better suited to plan observations and understand the data that will come from them. One can also find the structure of a source by numerically modeling the shape of the autocorrelation.

## 2. ANALYSIS BASED ON INTENSITY AUTOCORRELATION

Borra (2008) gives a rigorous theoretical analysis based on wave interaction effects but it may be difficult to follow for most readers. We shall now use an analysis based on intensity autocorrelation that is much easier to follow. It is based on the experimental results obtained by Borra (2011). Note that the intensity autocorrelation technique applies to interferometry with separate telescopes and also to gravitational lensing, because gravitational lensing causes interference effects that can be modelled with a Young interferometer; however, the discussion in this section is centered on interferometry using separate telescopes because it is easier to follow.

Borra (2011) used a very easy and practical way to obtain the spectrum of the current fluctuations that consists in digitizing the output current *I(t)* of the quadratic detector that measures the intensity signal from the interferometer and then performing with software the autocorrelation

$$I \otimes I = \int_{-\infty}^{\infty} I(t+t')I(t)dt \quad . \tag{1}$$

The power spectrum can then be obtained by taking the Fourier transform of the autocorrelation given by equation (1). This procedure, justified by the Wiener-Klintchine theorem (Klein & Furtak 1986), was used by Borra (2011) to obtain the power spectrum in the experiments that validate the theoretical work in Borra (2008). The analysis that follows is based on equation (1).

All of the relevant information about the angular intensity distribution of a source is contained in the autocorrelation. This can be understood by considering that the autocorrelation (equation 1) performs an integral over the product of the intensity *I(t)* measured by a quadratic detector as a function of time and *I(t+t')*. The intensity *I(t)* contains intensity fluctuations, caused by electromagnetic wave interaction effects, which vary over extremely short time scales. Klein & Furtak (1986) gives a convenient brief description of electromagnetic wave interaction effects that cause intensity fluctuations. Klein & Furtak (1986) also discusses the effect of the quadratic detector and the electronics, which average the fluctuations over long times. They smooth the fluctuations. The fluctuations have random amplitudes and are separated by times that vary at random. However, in the case of a point-like object that is observed by an interferometer, where the beam is split in two components that are then recombined by hardware or software, the recombined beam carries twin random fluctuations that are separated by a constant time $\tau$ due to the optical path difference (*OPD = c$\tau$*). We shall call the first fluctuation preceding fluctuation and call following fluctuation its twin that follows it after a time $\tau$. Let us consider the integral in equation (1) which gives a function of *t'*, the time by which the intensities are numerically displaced. As *t'* starts from a negative value and increases, in the autocorrelation integral in equation (1), the product *I(t+t') I(t)* gives an average value that varies linearly with *t'*, because every random fluctuation in *I(t)* multiplies another random fluctuation in *I(t+t')*, and so does the integral. However, as *t'* increases and approaches *t'* =- $\tau$, each preceding fluctuation in *I(t+t')* meets its twin contained in the following fluctuation in *I(t)*.thereby increasing the value of the integral in equation (1) and generating a peak in the autocorrelation. As *t'* keeps increasing the twin fluctuations no longer meet and the autocorrelation falls back to its average value. As *t'* further increases it then gives a second much stronger peak at *t' = 0* , where all fluctuations meet, and finally a peak at *t' = +$\tau$* where the preceding and following fluctuations meet again. This is the autocorrelation OPD dependence seen in figure 2 in Borra (2011) which shows peaks at OPDs at +- *6.35-m* and +- *3.12-m*. Note that peaks are present at two different OPDs in Borra (2011) because the experiment uses the autocorrelation of the combined intensities from two interferometers with two different OPDs (*6.35-m* and *3.12-m*). Note also that the average value of the autocorrelation is subtracted in the figure 2 in Borra (2011). The analysis in the present paragraph is therefore confirmed by the experiments in Borra (2011). The next two paragraphs extend this discussion to the detection of a binary star.

Let us now consider two telescopes separated by a very long baseline *B*. Figure 1 shows a layout of this two-element interferometer. For a point-like astronomical source they will detect the same intensity fluctuations at two different times separated by

$$\tau_B = \mathbf{B}.\mathbf{s}/c, \qquad (2)$$

where *B* is the baseline vector, *s* the unit vector in the direction of the source and *c* is the speed of light. A point-like source gives an electric field *E(t)* having an amplitude that fluctuates at random in time because of wave interactions. It is detected as identical electric fields that fluctuate at random but are delayed by the same time $\tau_B$ so that the first telescope sees $E_1(t) = E(t)$ and the second one sees $E_2(t)= E(t+\tau_B)$. The combined electric field from the two telescopes therefore contains identical twin random fluctuations caused by wave interactions that are separated in time by $\tau_B$. A quadratic detector then outputs an intensity signal, proportional to the time average of the square of the electric field, that also contains twin random intensity fluctuations separated in time by $\tau_B$ . Consequently the autocorrelation of the intensity as a function of time from the combined beams gives two peaks separated by *2 $\tau_B$* (one at -$\tau_B$ and the other at +$\tau_B$ ). Let us now consider a simple source, a binary star, made of two distinct point-like sources, *i* and *j*, separated by an angular separation *$\Delta\theta$* in the direction parallel to the baseline. The first telescope sees fluctuations separated by $\tau_B$ while the second sees a different time separation $\tau_B$ +$\Delta\tau$ with

$$\Delta\tau = (\mathbf{B}.\mathbf{s}_i - \mathbf{B}.\mathbf{s}_j)/c, \qquad (3)$$

where *B* the baseline vector and $s_i$ and $s_j$ are the two unit vectors pointing in the direction of the two sources. Equation (3) is the same relation used for aperture synthesis interferometry and is obtained from equation (2) for two different positions in the sky identified by the two unit vectors $s_i$ and $s_j$. In Figure 1, the unit vector $s_i$ would be the unit vector *s* and the unit vector $s_j$ a vector in a direction at a small angle with respect to the unit vector *s*. Equation (1) will therefore give two peaks separated by *2 $\tau_B$* for the autocorrelation of the contribution from the first telescope and two peaks separated by $2(\tau_B+\Delta\tau)$ for the contribution from the second one. This is a signal identical to the one contained in figure 2 in Borra (2011) which shows the autocorrelation of intensity from two different sources sending light through two different optically combined interferometers having different OPDs respectively of *3.12* and *6.35* meters. The difference between the two OPDs is thus *3.23* meters. This OPD difference will be used in the next paragraph to model a binary star observed by two distant telescopes.

Let us now consider the simple case of two stars separated by 1 arcsecond observed at a distance from the zenith of 45 degrees in the direction of the baseline vector ***B*** by two telescopes separated by 1000 kilometers. Prima facie a 1000 km separation sounds absurdly large; however as elaborated in section 6c in Borra (2013) which discusses optical telescopes, the technique allows such large separations. It also allows inexpensive low optical quality telescopes. It could therefore use two classical telescopes, e.g. one in Hawaii and another one in California, or two distant dedicated Cherenkov telescopes. Furthermore this baseline problem is far less important at longer wavelengths (e.g. the radio and infrared regions) and, in the optical region, will be less important with future technological improvements (Borra 2013). Using equation (3) we find that the optical path differences would be *3.43* meters, very close to the optical path difference between the two OPDs (*3.23* meters) in figure 2 in Borra (2011) (see previous paragraph). Consequently we can see that the separation as well as the relative magnitudes (obtained from the peak intensities in the autocorrelation) between the two stars of a binary can be directly obtained from the autocorrelation. Considering that the half-width of the autocorrelation peaks in Borra (2011) is less than *1/20* of the OPD separation, we can see that angular separations of the order of 0.01 arcseconds could be obtained with two telescopes separated by 1000 km. Furthermore, because the half-width of the autocorrelation peak is proportional to the inverse of the electrical bandwidth of the detector, even smaller separations could be measured by using a detector with a larger bandwidth than the one used by Borra (2011). The orientation of the axis of a binary system could be obtained by repeated observations with two telescopes having the same constant baseline separation but different angular orientation (that can come from the rotation of the Earth) with respect to the major axis of the binary system since the separation in the peaks of the autocorrelation will have a maximum value when the baseline is parallel to the axis and will be zero when it is perpendicular. This reasoning can be readily extended to another simple case where the angular intensity distribution follows a line, since it can be thought as a collection of stars of varying intensities aligned along a line. We therefore see that the autocorrelation function will give the needed information to obtain the angular intensity distribution.

This analysis can be extended to two-dimensional angular distributions by considering that the intensity of every sub-region of the source only autocorrelates with itself. Consequently one must simply add the autocorrelations of the intensities from the sub-regions. This can be understood from the rigorous mathematical analysis in Borra (2008) that shows that a microscopic source can only beat with itself. It can also be understood from the analysis based on autocorrelation that follows in the next sentences. The experiment in Borra (2011) uses two independent sources so that each generates independent electric field fluctuations that are separated by times that vary at random. Consequently the sum of the electric fields from the two sources, measured by the quadratic detector, has the sum of independent fluctuations that mix at random and is quite chaotic as a function of time. However the intensity autocorrelation in figure 2 in Borra (2011) shows that the autocorrelation generates two clearly separated peaks, in full agreement with the theoretical analysis in Borra (2008) that shows that each distinct microscopic source can only beat with itself, while the sources cannot beat among themselves, to generate a beat spectrum. The beat spectrum is the frequency spectrum of the intensity fluctuations *I(t)* given by the quadratic detector. Adding a large number of independent sources (say 10 for discussion purposes) would simply increase the number of distinct sources that add up from 2 to 10. Considering that the autocorrelation generates two distinct peaks from the added beams from two distinct sources, there is no reason why it would not generate distinct peaks from more sources. Consequently adding a large number of sources will give the sum of the autocorrelation peaks of each source.

For a source having a simple two-dimensional intensity distribution, the intensity distribution could be obtained by numerical modeling that would add the contributions of the autocorrelation peaks of every sub-region. It could also be obtained using the Fourier transform of the autocorrelation (Borra 2013). Note that this modeling can be done for a single constant baseline.

The fact that the sources in the experiment in Borra (2011) use erbium-doped fiber amplifiers, while stars are thermal sources, does not impact the analogy between a binary star and the experiment. This can be understood from the analysis based on intensity correlation in this section, which only uses the information contained in intensity fluctuations. Thermal sources have intensity fluctuations as well as non-thermal sources (Klein & Furtak 1986). The inconvenience that thermal sources have in the optical region comes from the fact that they have a small coherence time interval and consequently one need very strong sources. This is why non-thermal sources were used in Borra (2011). Note also that using separate non-thermal sources, if anything, can only help interference, while the purpose of the experiment was to show that the separate sources do not interfere,

## 3. COMPARISON WITH INTENSITY INTERFEROMETRY

A comparison between the technique and intensity interferometry can be made by using the discussion in section 2 which is based on the autocorrelation of intensity signals (equation 1). The physics of the proposed interferometry is a wave interaction phenomenon similar to the physics of the Hanbury-Brown Twiss intensity interferometer (Mandel 1962). It has similarities but also a major difference. While both intensity interferometry and spectral interferometry use the signals contained in intensity fluctuations, the outstanding difference comes from the fact that intensity interferometry uses the *cross-correlation of two different intensity signals* from two different telescopes, while spectral interferometry uses *the autocorrelation of a single intensity signal coming from the added beams* from two different telescopes.

The differentiation based on the autocorrelation and the cross-correlation explains why we have no limit to the maximum extendedness of a source that can be observed with the angular resolution set by a constant baseline, while intensity interferometry has an upper limit. In intensity interferometry the spatial coherence function, which quantifies the visibility of the interferometric signal, is obtained from the cross-correlation of the intensity signals from separate telescopes. Because the intensity fluctuation signals from an unresolved source seen by two different telescopes are exactly the same, the cross-correlation will give a strong signal at *t' =0*. On the other hand, as the extendedness of the source increases, one must add uncorrelated intensity contributions from different regions of the source. This rapidly decreases the cross-correlation signal. Let us now consider the autocorrelation used in this article. Because we autocorrelate as a function of *t'* (equation 1), the autocorrelation of an unresolved source will give a peak at *t'= $\tau_B$* that depends on the baseline *B* and the direction of the source (equation 2), another one at *t'= 0* and a third one at *t'= - $\tau_B$*. This kind of signal is shown in figure 2 of Borra (2011). For an extended source every intensity contribution of the different regions will autocorrelate with itself and give a peak at *t'=0*, and also give two peaks at different *t'=+- ( $\tau_B$ + $\Delta\tau$ )*, with *$\Delta\tau$* varying with the location of the individual separate regions. Figure 2 of Borra (2011) illustrates the situation for the simple case of two different OPDs, which is the kind of signal that a binary star would give (see the discussion in section 2). Their separate peaks can clearly be seen in the figure. Increasing the angular separation would simply further separate the peaks, so that they obviously would still be detected at very large separations. We can model a very extended source by the sum of a large number of star-like sources extending over its large angular diameter. The autocorrelation of the combined *I(t)* would give us the sum of peaks having the same width at different values of *$\Delta\tau$*. Consequently we see that there is no limit to the maximum angular diameter within which one could obtain the angular intensity distribution with the angular resolution set by the baseline. Note also that this modeling is done at a single baseline.

Because we consider wave interaction effects like intensity interferometry, we can also estimate the signal to noise ratio by applying similar considerations. Consequently, like in intensity interferometry, it is easier to work in the radio region than in the optical. This is discussed at length in Borra (2008) and Borra (2013) that show that this is a major inconvenient of the technique in the optical-infrared region but a minor one in the radio region. Like for intensity interferometry atmospheric seeing effects are not important in our case because the frequency of the current fluctuations at which the data are analyzed is much lower than the frequency of observation (Borra 2013).

## 4. OBSERVATION OF GRAVITATIONAL LENSES

Section 2 discusses very long baseline interferometry. We shall now briefly discuss the application of the autocorrelation to the observation of gravitational lenses. Gravitational lensing is a complex issue, as can be seen in a review paper by Treu (2010) and we obviously cannot carry out in here a detailed analysis. A gravitational lens takes the point-like image of a distant object and can transform it into a ring, multiple point-like images (e.g. the Einstein cross) or simply two point-like images. Two point-like images are more often observed and can conveniently be modeled by a Young interferometer (Refsdal 1964, Press & Gunn 1973). We shall use this case in the discussion below.

Considerable information on the source, the lens, the geometry of the event and the parameters of the universe can be obtained from the separation between the images, the relative brightness of the images and the time delays between the light beams of the images. The time delays are particularly important and relevant to the proposed technique. Presently time delays can only be obtained for strong gravitational lensing and require a source that has a luminosity that varies in time. In practice, this requires lenses having the mass of a galaxy that generate time delays greater than several days. On the other hand, the present technique works for a source that has a constant luminosity since it uses intensity fluctuations generated by wave interaction effects, which are present even in constant luminosity objects, and can detect large time delays as well as time delays as small as a few times the coherence length of the beam. Consequently time delays of the order of $10^{-7}$ seconds could be measured in the radio region and considerably smaller ones in the infrared and optical regions. As we shall see, there is no upper limit to the time delays that can be obtained. To obtain order of magnitude estimates of the time delay as a function of the mass of the lens, we can use the approximate relation

$$\tau = 3.4 \; 10^{-5} \, M, \qquad (4)$$

that Borra (2008) derived from Press & Gunn (1973), where the time delay $\tau$ is in seconds and $M$ is the mass in solar mass units. The parameters and approximations used to derive equation (4) are described in Press & Gunn (1973) who use average cosmological parameters (e.g. $z_{source} = 2$, $H0 = 60 \, km \, s^{-1} \, Mpc^{-1}$). Equation (4) can therefore give useful estimates for the present discussion. It shows that the $10^{-7}$ seconds delay from observations in the radio region would allow us to detect masses of the order of $3 \; 10^{-3}$ solar masses and that the smaller values of $\tau$ in the infrared or optical region would allow detections of even smaller masses. On the other hand, the upper limit to the time delays that can be measured is set by the time over which the data is taken, which gives the actual time limits in Equation (1) so that much larger masses could be measured. For example, 1 hour of observations would allow the detection of peaks separated in the range from $10^{-7}$ seconds and $3.6 \; 10^{3}$ seconds and therefore masses ranging between $10^{-2}$ and $10^{8}$ solar masses.

   The time delay $\tau$ can easily be obtained from the autocorrelation. To see this, let us model the gravitational lens using the simple Young interferometer model from Refsdal (1964) and Press & Gunn (1973) where the impact parameter $b$ (Press & Gunn 1973) is equivalent to the separation between the slits in a Young interferometer. The impact parameter $b$ depends on the mass of the lens as well as the cosmological parameters and the relative angular position between the source and the lens (Press & Gunn 1973). Considering this model, we can see that the lensing event generates two separate beams that have a time delay $\tau$ between them that depends on the impact parameter $b$ and therefore the mass of the lens. The detector of a telescope that observes a lensed source measures the intensity as a function of time from the two combined beams separated by a time delay $\tau$. Consequently, the autocorrelation will give a signal similar to one of the twin peaks seen in figure 2 in Borra (2011) which shows 2 twin peaks generated by the autocorrelations of two different pairs of combined beams respectively separated by optical path differences of *6.35-m* for one combined beam and *3.12-m* for the second combined beam. The twin peaks at +-*6.35-m* correspond to a time delay $\tau = 2.1 \; 10^{-8}$ seconds and a mass of *6 $10^{-4}$* solar masses using equation (4). A time delay $\tau = 2.1 \; 10^{-9}$ seconds (mass of *6 $10^{-5}$* solar masses) would give an OPD of 0.635-*m* that could easily be found by visual inspection. Even smaller time delays could be found with appropriate numerical techniques. We can see that one could obtain the mass of the lens more easily than with the technique described in Borra (2008) that necessitates a Fourier transform, since a simple visual inspection of the autocorrelation immediately gives the time delay. Because the time delay depends on the angular separation between source and lens, one could easily find variation of the separation by observing the objects at different epochs because the peak would change position in figure 2 in Borra (2011). A shorter time delay would displace it towards a lower OPD. In a two-beams gravitational lens the two beams may not have the same intensity but this will only decrease the strength of the peaks and not the delay time. The structure of the source could also be obtained from numerical modeling since a source can be modeled by point-like sources having different angular separations. Different angular separations give different time delays and one could therefore add peaks at different time delay $\tau$.

Time delay variations over relatively short times (e.g. hours or days) could also be detected within the large time delays (e.g. months) from strong lensing by autocorrelating the intensity fluctuations measured at different times separated by hours or days. One could therefore measure how the time delay $\tau$ varies with time in presently known strong gravitational lenses. This would give useful information. Note that it is not necessary to continuously observe the sources if one already knows the approximate time delay (e, g. a year) from previous observations of luminosity fluctuations. One could observe over a relatively short time, centered on the approximate time, numerically add the long time delay and perform the autocorrelation. Information on the angular intensity distribution of the source could also be obtained by using the simple modeling technique described in section 2.

When considering the observations of strongly lensed objects in the previous paragraph, one must however remember that the technique requires bright sources in the visible-infrared region (see last paragraph in section 3). Because the known strongly-lensed objects are distant quasars one would therefore have to observe in the far-infrared or radio region.

## 5. APPLICATION TO QUASAR MICROLENSING AND MILLILENSING

As an example of the application of the autocorrelation technique, we shall consider its use for the study of microlensing and millilensing in Quasars. Schmidt & Wambsganss (2010) give a review of Quasar microlensing and millilensing. They are caused by objects along the line of sight to the background quasars. They can be used to find the existence and effects of objects between the observer and the source. The objects can be stars or exotic objects (e.g. black holes). Lensing effects on quasars by compact objects having masses range between $10^{-6}$ and $10^3$ solar masses are called microlensing , while lensing effect by objects having masses ranging between $10^3$ and $10^7$ solar masses are called millilensing. Millilensing is particularly interesting because it could help understanding the nature of the dark matter in the halos of galaxies and also give information on the angular structure of the Quasar. Millilensing could be due to exotic objects such as subhalos made of dark matter or black holes having masses of several million solar masses .

Microlensing and millilensing (Schmidt & Wambsganss 2010) are conveniently observed in the individual images of strongly lensed Quasars. They are caused by objects in the Galaxies, or clusters of Galaxies, responsible for the macrolensing. Presently, microlensing is studied by observing the light curves of the individual images of strongly lensed Quasars (Schmidt & Wambsganss 2010). However, microlensing induces light curves that vary over mass-dependent time scales ranging from weeks to months and therefore require multiple observations over several months or years. Millilensing causes light variations over time scales ranging from several months to years and therefore require observations over extremely long times. Furthermore, light variations in the brightness of a quasar, besides being lens-induced, can also be intrinsic to the quasar. To distinguish between the two requires observations over times comparable to the time delays between the individual images of a macro-lensed Quasar, which are typically of the order of a year. Consequently, we can see that a large amount of effort and telescope time is required to observe millilensing and microlensing in Quasars from light curves. On the other hand, with the autocorrelation technique, a single observation over a short time (e.g. 10 minutes) in the radio region will unambiguously find the effect of microlensing and millilensing. The next few paragraphs elaborate on this.

A detailed analysis of microlensing and millilensing is complex (Schmidt & Wambsganss 2010) and would require a lengthy analysis that uses numerical simulations. This may be adequate for an article that analysis actual data but is beyond the scope of the present article, which concerns the introduction of the autocorrelation technique. In what follow we will therefore use the simple Young interferometer model which, although not rigorous, gives gives useful order of magnitude estimates.

Because the technique requires bright sources in the visible-infrared region and the known strongly-lensed objects are distant quasars, one has to observe in the radio region. Presently, Very Long Baseline Interferometry (VLBI) observations are carried out in the radio region. The electric field as a function of time $E(t)$ is recorded at each telescope, digitized and then sent to a digital correlator that processes the data for classical amplitude interferometry. Exactly the same data can be used for our purpose. One can take the digitized $E(t)$ and simply obtain the $I(t)$ needed in Equation 1 from $I(t) = <E(t)^2>$, where the time average $<>$ is carried out over times significantly larger (e.g. 100 times) than the sampling times. The VLBI electric field $E(t)$ varies too quickly to be sampled near its frequency of observation with current technology and is, instead, sampled at a lower beat frequency with a longer sampling time. Presently, at the NRAO observatory, VLBI observations at a frequency of *8.4 GHz* are obtained with a sampling time of *15.625* nanoseconds. A moving time average $<>$ over 100 samples therefore gives a continuous smoothing over a time of *1.5625 $10^{-6}$* seconds and intensity fluctuations having half-widths of the order of $10^{-6}$ seconds. Lensing would therefore give peaks in the autocorrelation similar to those in figure 2 in Borra (2010) having half-widths of the order of *$10^{-6}$* seconds. We see that delay times as low as *$10^{-6}$* seconds could be detected. Using Equation 4, this gives masses of the order of *$10^{-1}$* solar masses. On the other hand, 1 hour of observations would allow detection of peaks separated in the range between *$10^{-6}$* seconds and *$10^3$* seconds and therefore masses ranging between *3 $10^{-2}$* and *3 $10^7$* solar masses.

The experiments in Borra (2010) show that the position of the autocorrelation peaks vary linearly with optical path difference. Consequently the location of a peak would gives the time delay and the information needed to obtain the mass of the lensing object. Microlensing and millilensing from multiple lenses in a component of a strongly lensed quasar could also be easily detected since they would give multiple peaks in the autocorrelation. They would look like a series of peaks like those in figure 2 in Borra(2011).

The application of the technique to VLBI data is particularly interesting for the detection of millilensing events. Presently millilensing cannot be detected with observations of light curves since the observations would have to be carried out over several years. Other techniques could be used but, in practice, they are difficult and have not been used much (Schmidt & Wambsganss 2010). Because the present *15.625* nanoseconds seconds sampling of the VLBI data sets a $10^{-6}$ seconds lower limit in the time delays, we only could detect objects in the upper end of the microlensing range. However, there would still be two major advantages that come from the fact that the event would unambiguously be a microlensing event and that a single observation over a short time is required. With future technological improvements lower sampling times will eventually be possible and allow detecting lower masses.

Finally, millilensing can be used to study the structure of the Quasar. In principle, this can be done with a variety of presently existing, but, in practice, it is difficult to do it (Schmidt & Wambsganss 2010). In our case, we could find the structure by numerical modelling.

## 6. CONCLUSION

Although a rigorous theoretical treatment of this novel interferometric technique was carried out by Borra (2008), it is difficult to understand for astronomers unfamiliar with statistical optics. The discussion in section 2 uses a treatment based on the autocorrelation of intensity fluctuations that is far easier to understand. Furthermore, even for astronomers familiar with statistical optics, the autocorrelation technique is far more intuitive and therefore more useful to understand the data than Fourier analysis techniques. The principles and validity of the technique can be intuitively understood by considering the modeling of a binary star in section 2 which is based on the laboratory experiments in Borra (2011). As discussed in section 2, this modeling can readily be extended to more complex two-dimensional angular intensity distributions.

The analysis in section 2 also shows that the technique has a fundamental similarity with intensity interferometry for it also uses the signals contained in intensity fluctuations; however it also has a major difference that is discussed in Section 3. It comes from the fact that intensity interferometry uses the cross-correlation of the separate intensities, as a function of time, from two different telescopes, while this technique uses the autocorrelation of the intensity, as a function of time, of the combined beams. In particular, this difference shows that the technique allows one to measure large angular separations as well as small ones with only 2 telescopes separated by a constant baseline. This would allow one to measure extended sources with a high angular resolution using a constant baseline. On the other hand intensity and amplitude interferometers require either to move the telescopes to change the baseline or necessitate several telescopes located at different constant baselines.

Both intensity interferometry and the present technique use the correlations of intensity fluctuations. Consequently, if intensity fluctuations are detectable in astronomical sources observed with an intensity interferometer, which uses an integral similar to the one in equation (1) but for the cross-correlation between the intensities measured at two different telescopes, they also should be detectable with the autocorrelation in equation (1).

The discussion in Borra (2008) assumes that the optical path differences and the intensity distribution of the source observed will be obtained by Fourier analysis. However, the discussed in section 2 shows that one could also simply use numerical modeling of the shape of the observed autocorrelation signal to find the intensity distribution of the source.

The discussion in section 2 concentrates on the application of the autocorrelation to astronomical interferometry. However, presently, the technique is mostly useful for the study of gravitational lenses. Section 4 shows how the autocorrelation can be used for to the detection and analysis of gravitational lenses since they produce interfering optical beams that are commonly modeled by a Young interferometer (Refsdal 1964, Press & Gunn 1973). Section 5 gives a practical example for the observation of microlensisng and millilensing effects. It shows that presently available VLBI radio-astronomical data can be used to observe microlensing and millilensing in macrolensed Quasars. This could give very useful information on the structure of the galaxies and clusters of Galaxies responsible for the macrolensing. In particular, millilensing observations may give information on the dark matter substructures in the lensing galaxies.

## ACKNOWLEDGEMENTS

This research has been supported by the Natural Sciences and Engineering Research Council of Canada.

## REFERENCES

Borra, E.F. 2008, MNRAS, 389, 364
Borra, E.F. 2011, MNRAS, 411, 1965
Borra, E.F. 2013, MNRAS, 436, 1096-1101
Hanbury-Brown, R.1968, Ann. Rev. Astr. Astrophys, 6,13
Klein, M. V., & Furtak, T. E. 1986, Optics (New York: John Wiley & Sons)
Mandel, L 1962, J.Opt.Soc.Am., 52, 1335.
Refsdal S. 1964a, MNRAS, 128, 23
Schmidt, R.W, Wambsganss, J., 2010, Gen Relativ Gravit 42, 2127
Treu, T. 2010, Ann. Rev. Astr. Astrophys, 48, 87
Press H. W., Gunn J. E. 1973, ApJ, 185, 397

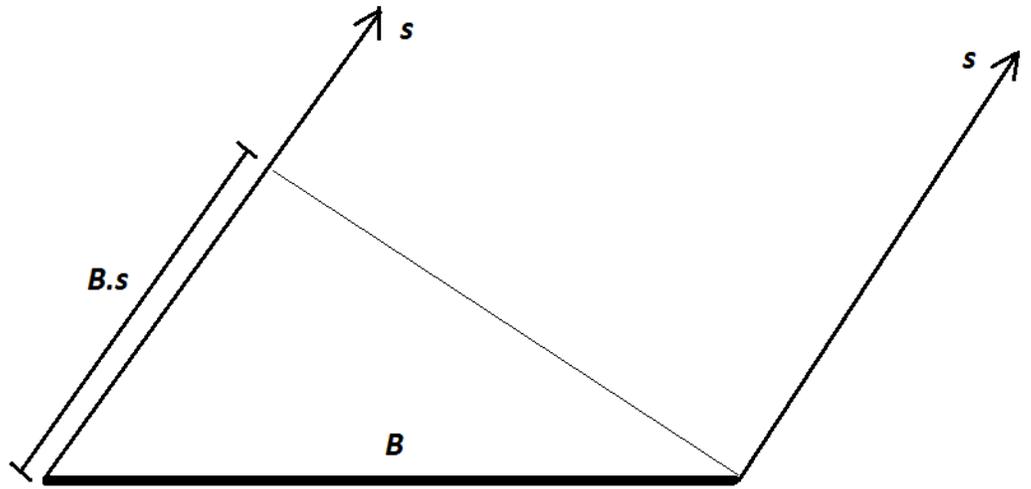

FIGURE CAPTIONS

**Figure 1**

It shows a layout of the two-element interferometer. The bold characters are used to indicate that these are vectors: **B** is the baseline vector (having the length of the baseline B), **s** is the unit vector in the direction of the source. The arrow-heads at the end of the lines show the directions of the vectors **B** and **s**. The vector product **B.s** gives the distance used to obtain the time delay $\tau_B = $ **B.s**$/c$ in Equation 2. A source having a complex structure could be modeled by adding several vectors pointing in directions slightly different than the single vector **s** in the figure.